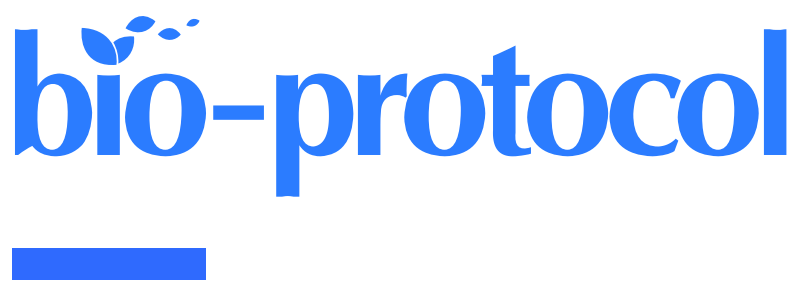

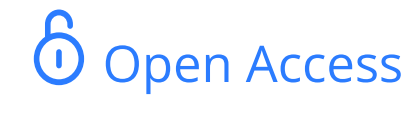

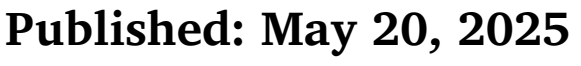

# From Bedside to Desktop: A Data Protocol for Normative Intracranial EEG and Abnormality Mapping

Heather Woodhouse[1], Sarah J. Gascoigne[1], Gerard Hall[1], Callum Simpson[1], Nathan Evans[1], Gabrielle M. Schroeder[1], Peter N. Taylor[1, 2, 3] and Yujiang Wang[1, 2, 3, *]

[1]CNNP Lab (www.cnnp-lab.com), Interdisciplinary Computing and Complex BioSystems Group School of Computing, Newcastle University, Newcastle upon Tyne, UK

[2]Faculty of Medical Sciences, Newcastle University, Newcastle upon Tyne, UK

[3]UCL Queen Square Institute of Neurology, Queen Square, London, UK

*For correspondence: yujiang.wang@newcastle.ac.uk

## Abstract

Normative mapping is a framework used to map population-level features of health-related variables. It is widely used in neuroscience research, but the literature lacks established protocols in modalities that do not support healthy control measurements, such as intracranial electroencephalograms (icEEG). An icEEG normative map would allow researchers to learn about population-level brain activity and enable the comparison of individual data against these norms to identify abnormalities. Currently, no standardised guide exists for transforming clinical data into a normative, regional icEEG map. Papers often cite different software and numerous articles to summarise the lengthy method, making it laborious for other researchers to understand or apply the process. Our protocol seeks to fill this gap by providing a dataflow guide and key decision points that summarise existing methods. This protocol was heavily used in published works from our own lab (twelve peer-reviewed journal publications). Briefly, we take as input the icEEG recordings and neuroimaging data from people with epilepsy who are undergoing evaluation for resective surgery. As final outputs, we obtain a normative icEEG map, comprising signal properties localised to brain regions. Optionally, we can also process new subjects through the same pipeline and obtain their z-scores (or centiles) in each brain region for abnormality detection and localisation. To date, a single, cohesive dataflow pipeline for generating normative icEEG maps, along with abnormality mapping, has not been created. We envisage that this dataflow guide will not only increase understanding and application of normative mapping methods but will also improve the consistency and quality of studies in the field.

## Key features

- Resultant normative maps can be used to test a broad range of hypotheses in the neuroscience field.
- Provides a more detailed walkthrough of the methods in the normative mapping study conducted by Taylor et al. [1] and other related publications [2–12].
- Offers flexibility: readers can tailor the final output by considering key decision points included throughout the protocol.
- Involves sub-pipelines, which may be useful to researchers in isolation (i.e., icEEG electrode localisation and/or interictal segment selection).

## Graphical overview

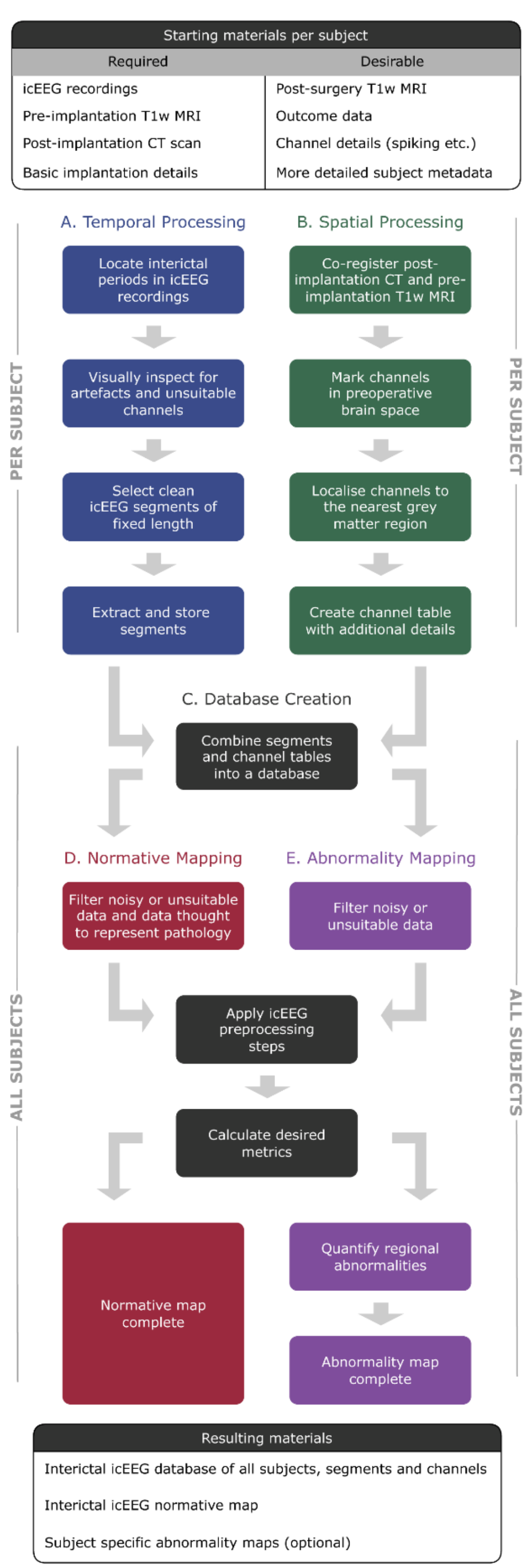

**Flowchart of pipelines involved in this protocol.** Temporal processing (pipeline A) and spatial processing (pipeline B) can be run sequentially or in parallel.

## Background

Normative mapping is a framework in which features of health-related variables are mapped at the population level. A common example is growth charts, used to evaluate whether a child's growth falls within the normal range for their age. In neuroscience research, this framework is widely used to map normative ranges of brain activity, for example.

However, normative mapping is particularly challenging in modalities that do not support measurements in healthy control subjects, such as the highly invasive intracranial electroencephalogram (icEEG). An icEEG normative map of brain activity is valuable because it allows us to better understand populations of interest. Further, individuals can be compared to the normative map, helping to identify, characterise, and localise potentially pathological abnormalities. Such work is available in the literature [1–7,13–20], revealing, for example, that icEEG normative mapping can localise epileptogenic tissue [1], and that the abnormality maps that follow are temporally stable [5].

Despite these promising findings, there is currently no detailed guide available for transforming clinical data, such as neuroimaging and icEEG recordings, into a normative map of brain activity localised to standardised brain regions. This literature gap hinders the consistent application of the methodology across studies. To remedy this, our protocol consolidates existing methods and provides key decision points for constructing normative maps comprising regional signal properties, along with (optionally) regional abnormality maps for new subjects.

One of the protocol's strengths lies in its modular design. Distinct sub-pipelines, such as the spatial processing pipeline, offer a standalone utility. For instance, our own lab has employed this pipeline to demonstrate that the incomplete resection of the icEEG seizure onset zone (SOZ) is not associated with post-surgical outcomes [10], amongst other results [8,9,11,12]. Furthermore, this protocol is written as a dataflow guide, allowing for flexibility. Researchers can choose their preferred programming language and data management software for implementation, decide whether to compute abnormalities, and apply the temporal processing pipeline to other modalities, such as scalp EEG (for instance). Our protocol allows for adaptation to best suit the reader.

While the protocol provides a standardised framework, certain aspects, such as manual resection mask delineation, could benefit from modernisation. Automated methods exist within our own lab [21] and externally [22–25], although their reliability varies [26]. The flexible, guide-like nature of our protocol is both a strength and a weakness. Importantly, we offer pseudo code and emphasise dataflow over low-level implementation, allowing researchers to tailor the process to their needs while fostering a deeper understanding of the method and related literature.

This protocol provides a cohesive data flow on icEEG normative and abnormality mapping, which is currently missing from the literature. We aim to equip neuroscience researchers with the tools needed to develop normative maps, allowing them to explore new hypotheses consistently and effectively.

## Equipment

1. PC or laptop capable of running the required software (below) with sufficient storage space. As a rule of thumb, we recommend at least 1 GB per subject. As an example, a standard Macbook Pro would be capable of running the full protocol. Increasing computational resources would allow a larger number of subjects to be processed more efficiently.

## Software and datasets

**Required**

1. icEEG recordings for each subject
2. Post-implantation computed tomography (CT) scan for each subject
3. Pre-implantation volumetric T1-weighted magnetic resonance imaging (T1w MRI) for each subject
4. Programming environment (this is the reader's choice, but common examples include MATLAB, Python, and R; see the General Notes section)
5. Software to view icEEG data (e.g., EDFBrowser or the reader's chosen programming environment)
6. FreeSurfer [27] (we used version 7.3)
7. Electrode localisation tools such as img_pip [28]
8. Database software (e.g., MongoDB)

**Optional/recommended**

9. Post-surgery T1w MRI for subjects who proceeded to resective epilepsy surgery
10. ANTS toolbox [29]
11. 3D image viewer such as FSLeyes or ITK-Snap
12. RAMPS pipeline [21]

*Notes:*

*1. If the reader is looking for suitable data, some of our published works [1–7] use the publicly available RAM dataset, found at [https://memory.psych.upenn.edu/RAM](https://memory.psych.upenn.edu/RAM).*

*2. Some of the above software may have certain CPU or license requirements. Please refer to each tool's documentation and/or official website page for the most up-to-date information.*

*3. We have included a glossary in the General notes section to avoid any ambiguity around our chosen vocabulary, e.g., electrode contact and electrode channel are often used interchangeably in the literature.*

**Prerequisites**

The experience of the reader should be high for programming, data organisation, neuroimaging, and EEG signal processing.

## Procedure

The folder structure throughout this protocol is the reader's preference. Figure 1 shows our recommendation for organising a subject's database and normative mapping files. Note that it is common to have data from multiple batches, in this case, hospitals. Throughout the protocol, we will implement unique subject identifiers, a combination of a four-letter abbreviation of the hospital and the subject's index within that hospital. For example, the first subject from University College London Hospital would have the identifier UCLH_001. This will be referred to throughout as the subject ID.

Our protocol comprises five sub-pipelines. Temporal processing (pipeline A) and spatial processing (pipeline B) can be run sequentially or in parallel. Both pipelines A and B are run on a per-subject basis and are hence repeated for each individual in the dataset. As such, in a cohort of 20 subjects, pipeline A would be repeated 20 times, as would pipeline B. The remainder of the protocol (pipelines C–E) is run with all subjects simultaneously. The outputs of pipeline A and B are combined to create a database (pipeline C). The database can be queried for subjects suitable for normative mapping (pipeline D). Abnormality mapping (pipeline E) is optional and typically applied to either new subjects or a subset of existing subjects that are held out following the stage of database creation (pipeline C).

```
Database-files/
├── UCLH_001/
│   └── exam/
│       ├── imaging/
│       │   └── 2009-05-23/
│       │       ├── pre_implantation_T1wMRI.nii
│       │       ├── post_implantation_CT.nii
│       │       ├── post_surgery_T1wMRI.nii
│       │       └── ...
│       └── icEEG/
│           └── 2010-11-28/
│               ├── icEEG_report.pdf
│               ├── eeg-segments/
│               │   ├── interictal_segment1.mat
│               │   ├── interictal_segment2.mat
│               │   └── interictal_segment3.mat
│               └── raw-eeg/
│                   ├── day1.edf
│                   ├── night1.edf
│                   └── ...
├── UCLH_002/
├── UCLH_003/
└── ...
Normative-mapping-files/
├── UCLH_001_icEEG2010-11-28/
│   ├── preprocessing/
│   │   ├── preprocessed_segment1.mat
│   │   ├── preprocessed_segment2.mat
│   │   └── preprocessed_segment3.mat
│   └── metrics/
│       ├── psd/
│       │   ├── psd_preprocessed_segment1.mat
│       │   ├── psd_preprocessed_segment2.mat
│       │   └── psd_preprocessed_segment3.mat
│       └── rbp/
│           ├── rbp_preprocessed_segment1.mat
│           ├── rbp_preprocessed_segment2.mat
│           └── rbp_preprocessed_segment3.mat
├── UCLH_002_icEEG2009-10-20/
├── UCLH_003_icEEG2012-01-30/
└── ...
```

**Figure 1. Example folder structure for database creation (pipeline C) and subsequent normative mapping (pipeline D)**

## A. Temporal processing: icEEG segment selection

The **goal of pipeline A** is to locate, extract, and systematically organise and store interictal segments from each subject.

Before starting this pipeline, some **decision points** (**DPs**) must be addressed. These are the reader's choice and will affect the resultant normative map. For each, we provide the decisions made in Taylor et al. [1] as an example.

• **Decision point A1:** What state of consciousness is the normative map? Taylor et al. [1] constructed a normative map of relaxed wakefulness. An alternative would be to construct a map of one of the standard sleep stages (N1, N2, N3, REM).

• **Decision point A2:** What constraints will be imposed regarding seizures to ensure selected data periods and segments are truly interictal? In Taylor et al. [1], interictal data must be a minimum of 2 h from any seizure activity. More detailed constraints are plausible, such as imposing a stricter constraint on the proximity of focal seizures compared to subclinical events.

• **Decision point A3:** Which sampling frequency will the interictal segments use to create the normative map? Taylor et al. [1] ensured all segments were at least sampled at 200 + Hz. Data with a higher sampling frequency than the intended level can still be used and downsampled at a later stage. However, icEEG data with a lower sampling frequency than the intended level should be excluded.

• **Decision point A4:** What is the desired referencing system? Taylor et al. [1] used the common average. This can still be changed later if data is saved in referential montage (as recorded), but the reader should have a default option. A bipolar montage is another common example.

• **Decision point A5:** What is the duration of the interictal segments used to construct the normative map? Segment length would ideally be consistent across all segments and subjects. Taylor et al. [1] used 70-s segments. However, the reader could also choose to use segments of differing length if it is absolutely unavoidable, and co-vary for the length later statistically.

• **Decision point A6:** How many interictal segments will be selected per subject? This will depend on how much data is available and may not be attainable for every subject. Multiple segments are selected to provide back-ups if a segment is excluded during preprocessing, and can also be used for later validation. Taylor et al. [1] selected three segments per subject where possible.

• **Decision point A7:** Will there be any time separation constraints on the selection of multiple segments? Taylor et al. [1] ensured that segments selected from the same subject and icEEG exam were separated by a minimum of 4 h.

Once **decision points A1–7** have been finalised, begin temporal processing (pipeline A). This is done on a per-subject basis and assumes subjects have labelled, long-term icEEG recordings comprising various wake states and periods of ictal and interictal activity. If the starting data is more tailored for the purpose, e.g., it is all interictal, some of steps A1–4 may be skipped. If at any step there is no data that meets the criteria, exclude this subject from further processing and begin with a new subject.

For a given subject...

1. Locate all periods within the icEEG files where the subject is in the required state of consciousness (**DP A1**) using the icEEG file labelling and annotations.

**Quality control:** A time constraint can be imposed to be more rigorous with the state of consciousness criteria. For example, only locate periods between 8 am and 10 pm when looking for awake recordings. Alternatively, apply a data-driven method of detecting sleep stages [30,31].

2. Discard any periods where the interictal constraints are not met (**DP A2**) using icEEG file labelling and

annotations. Figure 2 provides an example timeline of an icEEG exam for one subject. Following certain DPs being made, the figure demonstrates where suitable interictal periods would be located. Figure 3A provides an example of an ictal period, which must be avoided due to proximity to a seizure.

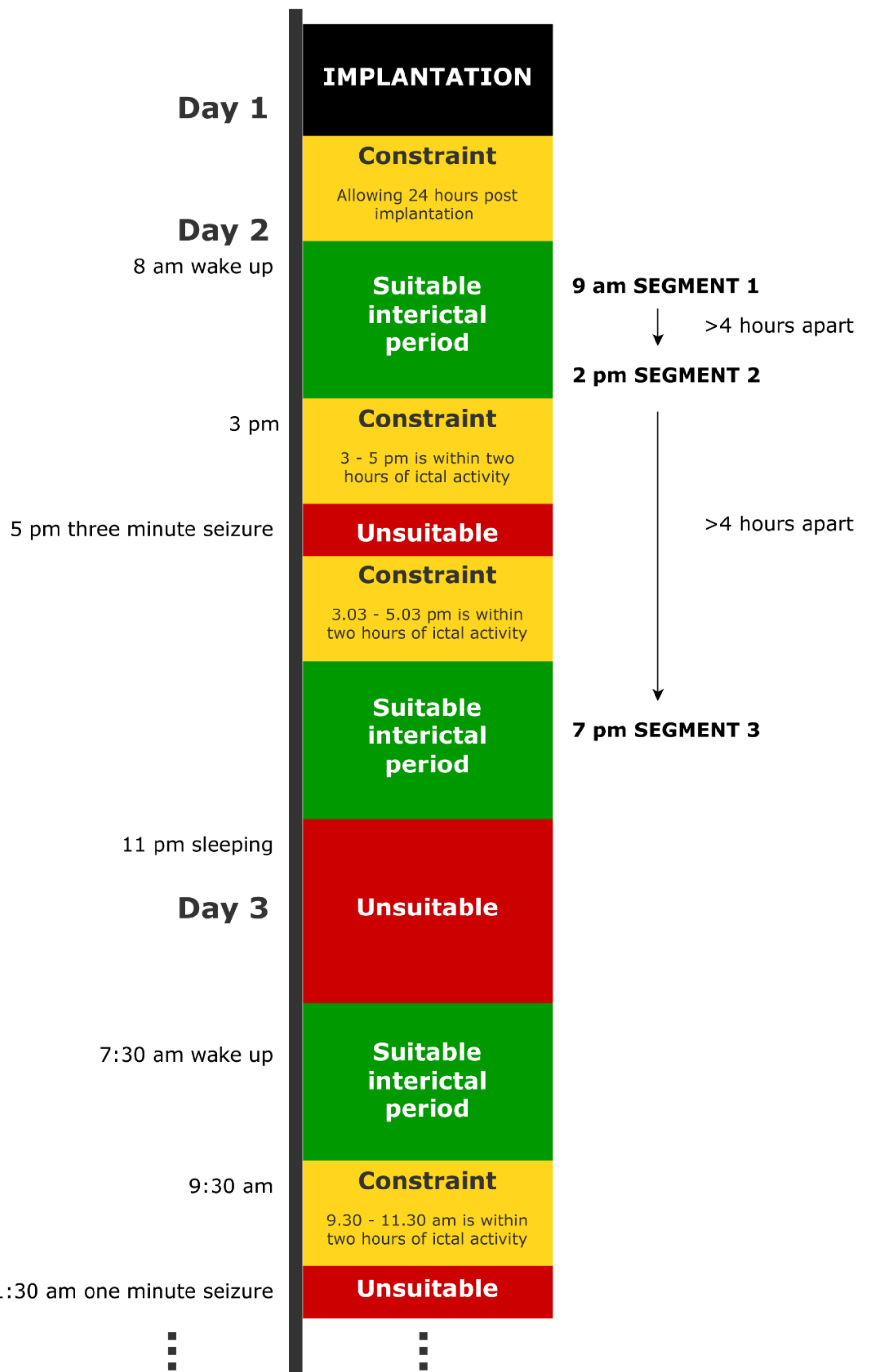

**Figure 2. Example timeline of an icEEG exam for a dummy subject, demonstrating where suitable periods of interictal data would be and the temporal location of the final interictal segments.** In this example, the normative map is being constructed in the wake state (**DP A1**), and three segments are being selected (**DP A6**) under the constraint that they are 2 h away from seizures (**DP A2**), 4 h away from one another (**DP A7**), and over 24 h after implantation (**quality control**).

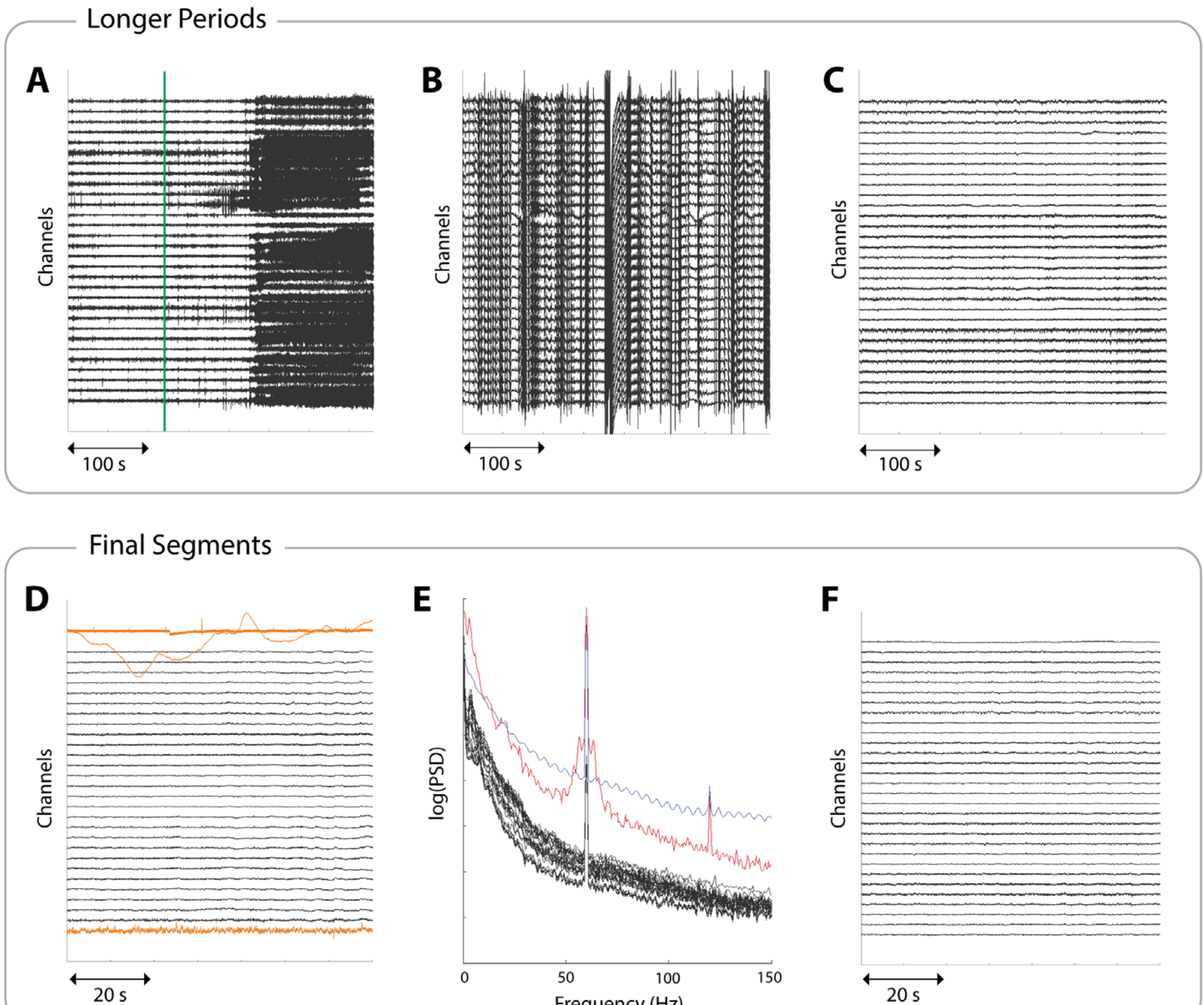

**Figure 3. Example icEEG data to aid in identifying suitable periods and segments. Each panel shows a different subject and a subset of 30 channels.** (A) A 380 s data period involving a seizure. The seizure onset, as marked by clinicians, is indicated by a green line. This line would correspond to the start of a red *unsuitable* box in Figure 2. (B) A 380 s data period showing non-physiological noise, indicated by the *streaky* appearance. Such large amplitudes are not representative of brain activity, and data exhibiting this should be discarded. (C) A clean 380 s data period, ready to proceed to segment selection. This would correspond to the *suitable interictal period* boxes in Figure 2. (D) A 70 s interictal segment with three examples of unsuitable channels highlighted in yellow. The segment is usable as long as these channels are noted so they can be removed downstream. (E) The power spectral densities (PSDs) of a 70 s interictal segment. Suitable channels are coloured black, showing a 1/f curve with power line noise at 60 Hz. The red line shows a channel with a broad peak around 60 Hz and an overall higher power in most frequencies (unlike the black channels), which most likely indicates a faulty channel. The blue line shows a channel that likely represents non-physiological activity due to its oscillatory pattern in the frequency domain. Both channels should be marked as unsuitable. (F) A clean 70 s interictal segment ready to be extracted and stored.

3. Discard any periods where the minimum sampling frequency is not met (**DP A3**).
4. Discard any periods that are shorter than the duration of the chosen segment length (**DP A5**).
5. Visually inspect the interictal periods for artefacts and noise and discard any noisy periods. Figure 3B demonstrates non-physiological noise, which may be present in some icEEG files. This can arise, for example, when channels temporarily disconnect, and should be avoided.

6. Retain only <u>clean</u> periods during the desired consciousness state. Figure 3C provides an example of a suitable, clean interictal period, which should be carried forward for further processing.

**Quality control:** If the subject has enough other periods available at this stage, avoid using the first 24 h (for example) post-implantation. This will minimise immediate implantation and anaesthesia effects. See Figure 2.

**Troubleshooting:** If data looks consistently noisy for a subject, before excluding this subject, we suggest checking if:

• Plotted amplitude is appropriate (i.e., when viewing many channel signals at once, a lower amplitude provides a clearer view).

• Necessary filters have been applied (for example, a low-pass filter if there is high-frequency noise, or a high-pass filter if there is suspected slow-wave activity or a current component).

• Desired re-referencing has been applied for visual inspection (**DP A4**; for example, common average or bipolar reference).

• Power line noise has been removed (note that power line noise can differ by country. Data originating from the UK, for example, may look cleaner after applying a 50 Hz notch filter, but data from the USA would require a notch filter at 60 Hz).

If the data still looks noisy, it likely needs to be excluded. If the aforementioned filters are unfamiliar to the reader, Sen et al. and De Cheveigne et al. provide discussions on EEG signal processing [32,33].

**Pause point:** At this stage, the reader should have identified periods of clean, interictal data for the subject. How many periods are available will vary per subject, as it is impacted by factors such as the number of seizures they had during the icEEG examination and how long the examination was.

7. Select the final segments of fixed length (**DP A5**, **DP A6**) from the periods of interictal data identified for this subject, ensuring all the imposed constraints are met (**DP A7**). Readers should check the power spectral densities (PSDs) as well as the icEEG when choosing final segments. Figure 2 demonstrates where segments might be selected from within suitable periods.

**Quality control:** If a segment has a small portion of unsuitable channels, but is otherwise usable, make a note of the unsuitable channels; they can be removed further downstream. Unsuitable channels include those that are noisy, faulty, or showing non-biological recordings. Such channels can be identified through visual inspection of icEEG (see Figure 3D) or through visual inspections of PSDs (see Figure 3E). PSDs should follow a 1/f shape and may have a narrow spike indicating power line noise, which can be dealt with at a later stage. Channels exhibiting a different pattern should be marked as unsuitable. More rigorous checks are carried out later, but visual checks are still useful here. Figure 3F shows a clean, interictal segment.

**Quality control:** Some channels in the interictal segments may exhibit epileptic activity, e.g., spikes. If any spiking channels are identified during segment selection, either visually or using a spike detection algorithm [34,35], they should be noted as such so they can be excluded further downstream.

**Pause point:** At this stage, the reader should have identified a selection of clean, interictal segments for the subject.

8. Create a suitable folder to store this subject's extracted segments. See Figure 1 for our recommendation.

9. For one segment, use your preferred programming language to extract the following data from the original file format (e.g., extract the relevant data from a .edf file and store it as a .mat file). These should be stored within the folder created in the previous step.

a. All channel names

b. Names of channels labelled as unsuitable (a subset of point a.)

c. Names of channels labelled as spiking (a subset of point a.)

d. Time series data (a matrix of size channels × time points)
e. Date and time at the chosen precision (e.g., second) at the start of the segment
f. Duration of the segment
g. Sampling frequency
h. State of consciousness
i. Subject ID
j. Hospital at which the data was recorded
k. Segment number
10. Repeat step 9 for all remaining segments.

Repeat the 10 steps above for all subjects.

**Endpoint:** A selection of interictal, fixed-length icEEG segments in the desired state of consciousness extracted for each subject, with metadata included and unsuitable channels noted.

## B. Spatial processing: electrode localisation and resection mask delineation

The **goal of pipeline B** is to localise electrodes to standard brain regions of interest (ROIs, usually in cortical grey matter or amygdala/hippocampus in this work), record additional channel details, and, where applicable, create resection masks. Before starting this pipeline, some **decision points** (**DPs**) must be addressed. These are the reader's choice and will affect the resulting normative map. For each, we provide the decisions made in Taylor et al. [1] as an example.

• **Decision point B1:** What constraint will be imposed for excluding channels based on distance from grey matter? Taylor et al. [1] excluded channels that appear to be located more than 5 mm from grey matter.
• **Decision point B2 (optional, requires resection mask):** What threshold will be implemented to define a channel as recording from subsequently resected tissue? Taylor et al. [1] defined channels as recording from resected tissue if they were within 5 mm of the resection mask.

Once **decision points B1 and B2** have been finalised, begin spatial processing (pipeline B). This is done on a per-subject basis and assumes subjects have a pre-implantation T1w MRI, a post-implantation CT scan, and, if drawing resection masks, a post-surgery T1w MRI. Clinical reports may have information on which channels were recording from subsequently resected tissue. If so, **decision point B2**, step B5, and steps B7e–g can be considered unnecessary. Be aware that not all subjects proceed to resective epilepsy surgery. Subjects without resection masks or resection details may still be useful, depending on the subsequent research questions. If at any step there is no data that meets the criteria, exclude this subject from further processing and begin with a new subject.

For a given subject…

1. Run the FreeSurfer "recon-all" pipeline on the pre-implantation 3D T1-weighted MRI.
*Note: Following successful completion, the file mri/orig.mgz will be generated. This file has 1 mm isotropic voxels and is the reference space for all future analyses. This step also generates the file mri/aparc+aseg.mgz, which contains the volumetric parcellation of different ROIs. This uses the Desikan–Killiany parcellation [36] built into FreeSurfer and is*

*used later in step B7.*

**Quality control:** Check the outputs of step B1 and consider performing manual edits via control points where appropriate [37].

2. Co-register the post-implantation CT scan to the orig.mgz file. This may require conversion from mgz format to nifti (nii.gz) format. For this, the FreeSurfer mri_convert command can be used. Registration should have a rigid body (6 degrees of freedom), without deformation. The ANTS toolbox [29] contains a highly effective set of tools for this purpose. The following command can be used from the toolbox:

```
/usr/local/ANTs/bin/antsRegistration --verbose 1 --dimensionality 3 --float 0 --
collapse-output-transforms 1 --output [ <Output_FilePaths>] --interpolation Linear -
-use-histogram-matching 0 --winsorize-image-intensities [ 0.005,0.995 ] --initial-
moving-transform [ <input_orig.nii.gz>,<input_CT.nii.gz>, 1 ] --transform Rigid[ 0.1 ]
--metric MI[<input_orig.nii.gz>, <input_CT.nii.gz>,1,32,Regular,0.25 ] --convergence
[ 1000x500x250x100,1e-6,10 ] --shrink-factors 8x4x2x1 --smoothing-sigmas 3x2x1x0vox
```

The rigid registration uses mutual information (MI) to measure concordance between the images. Since the images are from different modalities, contrast for specific tissues will be altered. MI compares histograms of the two images and aligns them, detecting similar anatomical patterns between the images, even if they have contrasting intensities. This method is widely used for multi-modal registration.

**Quality control:** Following registration, visually inspect that the images are correctly aligned using software for viewing 3D imaging data, such as FSLeyes or ITK-Snap (see Figure 4A for an example of a correctly aligned image).

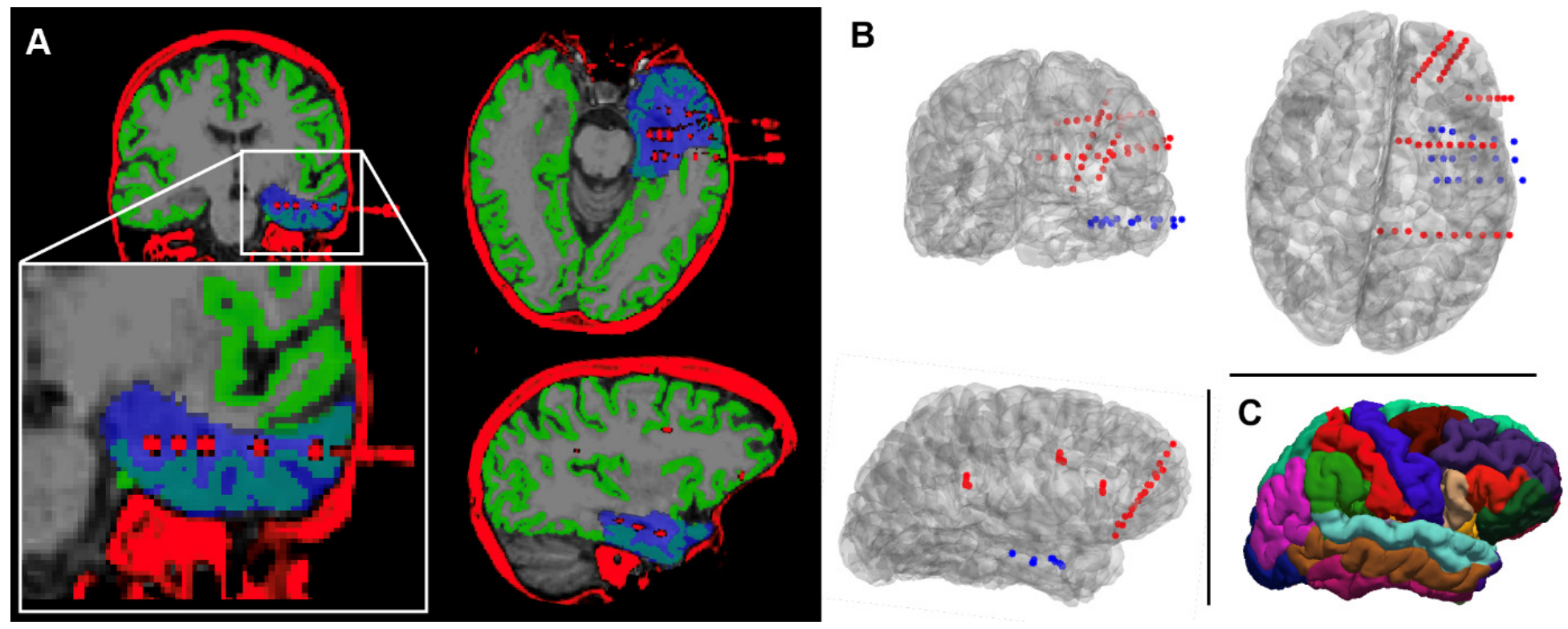

**Figure 4. Quality control checks and outputs from pipeline B for an example subject.** (A) Correctly aligned post-implantation CT scan and orig.mgz file following co-registration (**quality control**, step B2). Overlaid in blue is the resection mask, which delineates the tissue that was subsequently resected during surgery (**quality control**, step B5). The mask was generated using the RAMPS pipeline [21]. (B) Rendering of the subject's implanted electrodes following channel localisation (**quality control**, step B4). The channels on each electrode are indicated by coloured dots; blue channels were defined as recording from subsequently resected tissue, whilst red channels were spared (step B9). Looking at both panels A and B, it can be determined that the subject had surgery on the temporal lobe. (C) Visualisation of the Desikan–Killiany parcellation [36] to which channels will be localised.

3. Manually mark channels in pre-operative brain space using, for example, the open-source Python package

`img_pipe` [28].

*Note: A full protocol for this software has been previously published [28]. Here, we summarise the section "electrode identification on the co-registered CT scan" to obtain Cartesian coordinates for each channel.*

a. Follow the instructions in Hamilton et al. [28] to launch an interactive Python graphical user interface (GUI) showing the registered CT scan overlaid on the skull-stripped MRI.

b. Navigate the crosshairs to channel 1 of an electrode in the brain. Channel 1 is usually the deepest point of contact, but you should double-check the implantation information to confirm this.

c. Add a new electrode by pressing the *n* key, triggering a prompt to name it. For simplicity, we recommend following the clinician's naming systems. This is often anatomical abbreviations (HIPP, AMYG…) or simple alphabetical labelling (A, B…).

d. Press the *e* key to add a channel at the crosshair position. A coloured shape will appear in the GUI indicating where the channel has been marked, along with a legend showing the electrode name. The channel coordinates will be automatically saved in a folder named after the electrode.

**Quality control:** Ensure the crosshairs are in the centre of the channel in *all* views in the GUI (coronal, sagittal, axial).

e. Continue to mark the channels on the current electrode by pressing *e* over each one. Work in order up the electrode from the deepest to the most superficial channel.

f. Repeat steps B3b–e above for each electrode in the brain.

4. Once the channels are marked for all electrodes in the subject's brain, export the channel locations. If using the pipeline in Hamilton et al. [28], these will be in the same space as the orig.mgz file described in step B1.

**Quality control:** Write custom code or follow the guidelines in Hamilton et al. [28] to produce a rendering of the electrodes in the subject's brain space. Visually confirm that the electrodes are in the expected locations by comparison with implantation information. Figure 4B provides an example of this.

**Pause point:** At this stage, the reader should have a folder containing channel locations for each of the subject's implanted electrodes.

5. [Optional] Generate a resection mask in pre-implantation orig.mgz space. We recommend the RAMPS pipeline [21]. In brief, this resection mask pipeline generates a mask by:

a. Registering a post-surgery T1w MRI to the pre-implantation T1w MRI (using the ANTS toolbox [29]).

b. Performing tissue segmentation to identify a resection cavity (using ANTS ATROPOS methods [29]).

c. Performing multiple rounds of segmentation, dilation, and erosion to generate an accurate delineation.

Once installed, the RAMPS pipeline can be executed with the following command:

```
python /Path_to/RAMP.py </Path_to/Pre-implantation-Scan.nii.gz> </Path_to/Post-surgery-Scan.nii.gz> </Path_to_Output_Folder_file_path/> <Output_Prefix> <Hemisphere> <Lobe>
```

where `<Hemisphere>` is the hemisphere containing the resection (e.g., "L" or "R"), and `<Lobe>` is the lobe(s) of resection (e.g., "T" for temporal, or "F" for frontal).

**Quality control:** Overlay the resection mask with the orig.mgz file to ensure it is in the expected location. Figure 4A includes a visualisation of a resection mask.

6. Generate a table for each channel that contains the following information:

a. Name of the channel (electrode name, followed by channel number, e.g., HIPP1).

b. Electrode type (such as "G" for grid electrodes, "D" for depth electrodes).
c. Hemisphere from which each channel records ("Left", "Right").

7. Write code in the reader's preferred programming language to:
a. Load the channel locations (generated in step B3) and a table containing channel information (generated in step B6).
b. Load the aparc+aseg.mgz file (generated in step B1), which contains volumetric grey matter ROIs, and a volumetric segmentation of the white matter. A surface representation of the parcellation is shown in Figure 4C.
*Note: If the reader is using MATLAB, the FreeSurfer function `MRIread.m` can be used for this.*
c. Localise channels to the nearest grey matter region in the same hemisphere by minimising the Euclidean distance.
*Note: It is possible that a channel implanted between the two hemispheres may be closer to a region in the opposite hemisphere than that from which it records. It is therefore important to constrain by the known side.*
d. Exclude channels that are more than the specified distance from grey matter (**DP B1**)/are located in white matter by setting their assigned region to 'n/a'.
e. [Optional] Load the resection mask generated in step B5.
f. [Optional continued] Compute the Euclidean distance from each channel to each voxel located in the mask.
g. [Optional continued] Save the minimum distance computed in the previous step.
**Pause point:** At this point, the reader should have localised channels to ROIs, excluding those in white matter or located too far away from grey matter. Optionally, for each channel, the reader should have recorded the minimum distance of each channel to the resection mask.

8. Load the output from steps B6 (channel coordinates and names) and B7 (ROI localisations and distances from channels to resection masks) into your preferred programming software.

9. Create a final table encompassing all channel details for the subject, with the variables/columns outlined below. Table 1 provides an example channel table for one electrode. Table 2 shows the desirable (but not essential) variables for the same example electrode.
a. Channel name (e.g., HIPP1, first recorded in step B3).
b. Coordinates (x, y, z) of the channel (first recorded in step B3) in the same space as orig.mgz.
c. Type of electrode that the channel is on (step B6).
d. Hemisphere from which the channel records (step B6, "Left", "Right").
e. Numerical index of the ROI to which the channel has been assigned using the Desikan–Killiany atlas. A mapping from numerical index to region is available in Section 2 of a ROI tutorial on the FreeSurfer Wiki [38]. A visualisation of the atlas is provided in Figure 4C.
f. [Optional] Binary indicator of whether the channel was recording from tissue that was subsequently resected (using the output from step B7g) (**DP B2**).

The following variables are desirable, but not essential:
g. Binary indicator of whether the channel exhibited spikes in the interictal icEEG recordings (found in clinical reports if available or may have been identified in pipeline A).
h. Binary indicator of whether the channel is within any structural abnormality, such as a lesion (found in clinical reports if available or identified through visual inspection of neuroimaging data).
i. Binary indicator of whether the channel is located within the clinician-defined SOZ.

**Table 1. Channel table for one electrode targeting the hippocampus in an example subject**

| name | location orig | | | electrode type | hemisphere | ROI ID | ROI name |
|---|---|---|---|---|---|---|---|
| LHIPP1 | -29.83 | -15.10 | -9.69 | depth | left | 17 | Left-Hippocampus |
| LHIPP2 | -34.91 | -15.10 | -8.60 | depth | left | 17 | Left-Hippocampus |
| LHIPP3 | -39.70 | -15.10 | -7.65 | depth | left | 17 | Left-Hippocampus |
| LHIPP4 | -44.59 | -14.44 | -6.77 | depth | left | 39 | lh-middle-temporal |
| LHIPP5 | -49.29 | -14.44 | -5.68 | depth | left | 39 | lh-middle-temporal |
| LHIPP6 | -54.36 | -13.71 | -4.86 | depth | left | 39 | lh-middle-temporal |
| LHIPP7 | -58.97 | -13.71 | -4.29 | depth | left | 39 | lh-middle-temporal |
| LHIPP8 | -64.14 | -13.24 | -3.15 | depth | left | 39 | lh-middle-temporal |
| LHIPP9 | -68.00 | -13.24 | -2.46 | depth | left | 39 | lh-middle-temporal |
| LHIPP10 | -73.33 | -13.24 | -1.29 | depth | left | 39 | lh-middle-temporal |

**Table 2. Continuation of Table 1, showing an example of the additional, desirable variables that should be recorded in the channel tables, if available.**

| name | Channel recorded from tissue that was... | | | |
|---|---|---|---|---|
| | subsequently resected | exhibiting spikes | structurally abnormal | within the clinician-defined SOZ |
| LHIPP1 | 0 | 1 | 0 | 1 |
| LHIPP2 | 0 | 1 | 0 | 1 |
| LHIPP3 | 0 | 1 | 0 | 0 |
| LHIPP4 | 0 | 1 | 0 | 0 |
| LHIPP5 | 0 | 1 | 0 | 0 |
| LHIPP6 | 0 | 0 | 0 | 0 |
| LHIPP7 | 0 | 0 | 0 | 0 |
| LHIPP8 | 0 | 0 | 0 | 0 |
| LHIPP9 | 0 | 0 | 0 | 0 |
| LHIPP10 | 0 | 0 | 0 | 0 |

10. Save the resulting channel table for this subject using the subject identifier in the file name.

*Note: With the exception of the original neuroimaging files and the final channel table, the reader is free to discard all other tables, variables, and files created during this pipeline. The information in the final channel tables will ultimately be recorded in the database (created in pipeline C). Hence, channel tables are not included in the example folder structure in Figure 1.*

Repeat the 10 steps above for all subjects.

**Endpoint:** A channel details table for every subject, containing the name, location, and regional localisation of every channel the subject had implanted, along with other channel-level metadata where available.

## C. Database creation

The **goal of pipeline C** is to create an organised database that can be queried for downstream work.

Database creation aggregates all subjects' data and uses the outputs from temporal processing and spatial processing (pipelines A and B). This database will include all subjects' data and channel information (including channels marked as unsuitable, resected, etc.). The choice of database software is the reader's personal preference. In our case, we used MongoDB as a NoSQL database for increased flexibility and scalability in a range of research applications, not limited to normative mapping. In theory, however, for the purposes of this paper, relational databases or even using several spreadsheets could also be a suitable method of databasing. The user may wish to use relational databases if the database is limited in its purpose (e.g., only for normative mapping), and a stronger guarantee of data normalisation is required. We would not recommend using several spreadsheets if the database is highly complex with one-to-many or many-to-many relationships, e.g., where several subjects have repeat visits or second surgeries, or when several hospital sites and subjects are involved.

Instead of a step-by-step process, we here describe the structure of our database in detail, to allow readers to create similar structures using their preferred software. See Figure 5 for a visualisation of the database structure. At the end of this section, we also include the Object-Data-Map for MongoDB (written with Python and mongoengine syntax) for completeness and clarity.

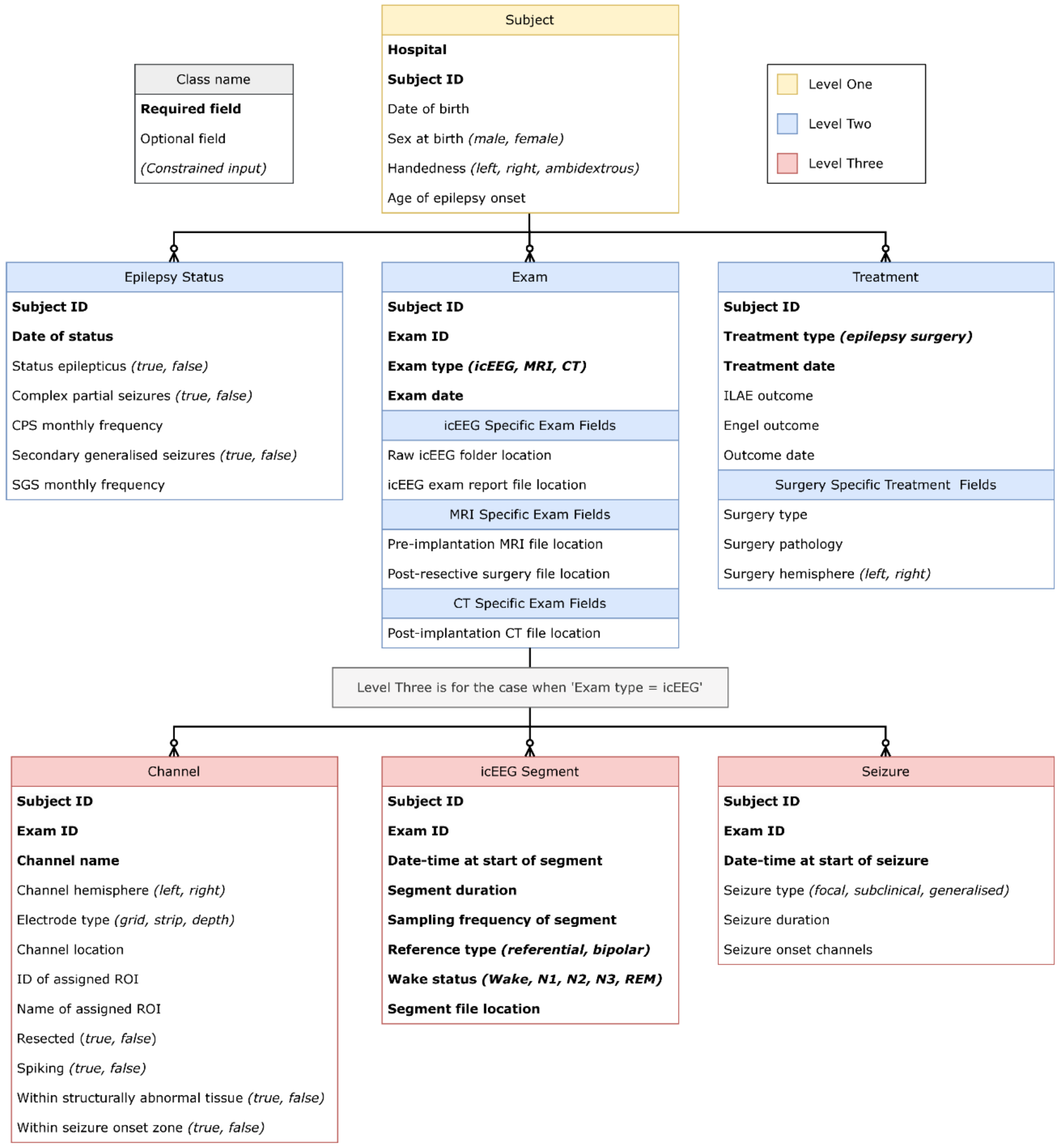

**Figure 5. Visualisation of the recommended database structure as described in detail throughout pipeline C.** All black lines indicate one-to-many relationships. The yellow box indicates the highest level of the database structure, Subject at level one. Subjects can have anywhere between 0 and 3 of the subsequent level-two classes, indicated by their blue colour (Epilepsy Status, Exam, Treatment). When a subject has an icEEG exam, there is also the option to input any (or none) of the three level-three classes, indicated in red. These are Channel, icEEG Segment, and Seizure. Within each class, required fields are indicated in bold font, whilst optional fields are in standard font. Constrained inputs are indicated by italics in parentheses.

To begin, we list some basic database requirements, which apply regardless of software choice:

- The database consists of various *classes,* which define its structure.
- Classes have properties, which we call *fields*, where we input information.
- Fields can be *required*, meaning we must have input to make an entry, or they can be *optional*.
- To ensure data quality, field inputs can be subject to certain *constraints*, meaning they can only take certain values.

Specifically, we have outlined the structure for our icEEG database below. As in Figure 5, required fields are **bold**, and constraints are indicated by *italics*. In the text, an underline is used to indicate a class. For clarity, database classes and fields are capitalised.

## Level one

### Level one: Subject

Subject is the highest-level class. We must have the Subject class to be able to input any lower-level data, such as the icEEG segment information. All lower-level classes are optional; we can simply record knowledge of an individual subject without any further specifics on their examinations or treatments.

Required fields in the Subject class are **Hospital** and **Subject ID**. Here, **Hospital** denotes the site at which the subject was recorded; we choose to constrain it to a 4-letter key, for example, ULCH for University College London Hospital. As mentioned previously, **Subject ID** combines the individual's **Hospital** code with a unique numerical identifier. The **Subject ID** appears in all subsequent classes; as such, we can follow each subject throughout the levels of the database.

Optional fields are reader's preference. We include Date of birth (or simply Year at birth), Sex at birth, Handedness, and Age of epilepsy onset. Sex at birth is an example of a field that should be constrained, taking exactly one entry out of *male* and *female*. Similarly, Handedness should be constrained to *left*, *right*, and *ambidextrous*.

## Level two

Level two comprises three classes that branch off from Subject (level one). These are Epilepsy Status, Exam, and Treatment. There is an optional, one-to-many relationship between Subject and each of these lower-level classes, as indicated by the connecting lines used in Figure 5. This means a given subject could have any number of entries for each level-two class (including no entry). Each level-two class is outlined in more details below.

**Level two: Epilepsy Status**

Epilepsy Status records information about the state of the individual's medical condition. This would, for example, reflect the epilepsy status at a particular point of follow-up with the subject, e.g., capturing if they had seizures in the recent past. A Subject may have multiple records of their epilepsy status across a few years, or they may have no record of epilepsy status at all.

Required fields in the Epilepsy Status class are **Subject ID** and **Date of status**. The **Date of status** field must be unique within the **Subject ID**.

Optional fields we include are Status epilepticus, Complex partial seizures, and Secondary generalised seizures, which all take a Boolean input of *true* if the individual had the seizure type, and *false* otherwise. Additional optional fields include the monthly frequency of each seizure type.

**Level two: Exam**

Exam refers to the assessments performed, e.g., to try and localise the seizure origin. In our case, individuals often have both icEEG and neuroimaging exams available. Further, a Subject may have more than one icEEG recording session. Each different exam requires a new entry.

Required fields in the Exam class are **Subject ID**, **Exam ID**, **Exam type**, and **Exam date**. Similarly to **Subject ID**, **Exam ID** is a string that is unique to each of the subject's exams. It appears in all subsequent classes (level three), allowing the reader to trace back to the specific exam. **Exam type** is constrained to a list of values, such as ***icEEG***, ***MRI***, or ***CT***.

For each **Exam Type**, there are further optional fields indicating the file location of the raw icEEG or neuroimaging files, along with any available related files such as the icEEG report. Note that raw data is not stored in the database itself; just the location is recorded. For example, taking Figure 1 as a suitable folder structure, for subject UCLH_001's icEEG exam, the database entry for Raw icEEG folder location would be "Database-files/UCLH_001/exam/icEEG/2010-11-28/raw-eeg".

**Level two: Treatment**

Treatment refers to the approach taken to manage epilepsy, such as surgery. Note that treatment data might not be available for some individuals. Alternatively, an individual may have two resective epilepsy surgeries and thus will have two Treatment entries.

Required fields in the Treatment class are **Subject ID**, **Treatment type**, and **Treatment date**. In our data, **Treatment type** is constrained to ***epilepsy surgery*,** but in theory, medication treatments can also be captured here**. Treatment date** must be unique for any combination of **Subject ID** and **Treatment type**.

Optional fields include Outcome fields (for ILAE, Engel, or both) and Outcome date. In the case where **Treatment type** is ***epilepsy surgery***, there are additional optional fields including Surgery type, Surgery pathology, and Surgery hemisphere. The latter is an example of a constrained field, taking one entry out of *left* and *right*.

Level three

Level three comprises three classes that branch off from Exam (level two) and only become available when the subject has an input for Exam and that input includes **Exam type** = ***icEEG***. The level-three classes are Channel, icEEG segment, and Seizure. If an individual had more than one icEEG exam, the subsequent level-three classes corresponding to each session will be distinguished using the **Exam ID** defined in level two.

There is an optional, one-to-many relationship between Exam and each of these lower-level classes, as indicated by the connecting lines used in Figure 5. This means that a particular subject's icEEG exam could have any number of entries for each level-three class (including no entry). Each level-three class is outlined in more detail below.

**Level three: Channel**

Channel refers to the recording channels on each electrode that the subject had implanted in the icEEG exam. Subjects can have any number of channels implanted, and each one has its own Channel input. This class is where all the information in the channel tables created in spatial processing (pipeline B) is stored. For a given subject, each row of their channel table would correspond to one Channel entry.

Required fields in Channel are **Subject ID**, **Exam ID**, and **Channel name**. The latter must be unique within this exam (using **Exam ID**) and is typically a combination of the electrode name and the channel's location on the electrode, e.g., in our data, HIPP1 would be the deepest channel on the electrode targeting the hippocampus.

Optional fields include Channel hemisphere, Electrode type, Channel location, ROI ID, and ROI name. There are a few constraints on these fields:
• Channel hemisphere should be constrained to *left* and *right.*
• Electrode type refers to the type of electrode the channel sits on and is constrained to *grid, strip,* or *depth.*
• The ROI is the brain parcellation that the channel is assigned to and is identified using both a number (ROI ID) and a label (ROI name).
• ROI ID and ROI name should be one of those found in the Desikan–Killiany parcellation [36].
There is a further set of optional, Boolean fields referring to the suitability of the channel for normative mapping. These are indicators of whether the channel is Resected, Spiking, Within structurally abnormal tissue, or Within the SOZ. Such information may be found in clinical reports or may be ascertained in previous pipelines.

**Level three: icEEG segment**

icEEG segment stores information on the interictal segments selected in pipeline A. Some subjects may not have any suitable interictal segments in their icEEG exam, whilst others may have had several selected (**DP A6**).
Required fields in icEEG segment are **Subject ID**, **Exam ID**, and the **Date-time**, **Duration**, **Sampling frequency**, **Reference type**, **Wake status**, and **File location** of the segment. In our data, the **Reference type** is constrained to ***referential*** and ***bipolar***. The segment's **Wake status** should be constrained to the standard sleep stages (***wake, N1, N2, N3, REM***), with ***NA*** as a final option for any segments where the state of consciousness could not be determined. Again, taking Figure 1 as a suitable folder structure, for subject UCLH_001's first interictal segment, the database entry for **Segment file location** would be "Database-files/UCLH_001/exam/icEEG/2010-11-28/eeg-segments/interictal_segment1.mat".

We do not include any optional fields in icEEG segment.

**Quality control:** As an extra precaution, decision points from temporal processing (pipeline A) can be quality control checked here. For example, do the inputted segments have the desired duration (**DP A5**)? Are multiple segments separated by the outlined time constraints (**DP A7**)?

**Level three: Seizure**

Seizure stores information on ictal events that may have occurred during the examination. Subjects may have any number of seizures during the icEEG exam (including none). Seizure data itself is not of interest for this pipeline; however, knowing what type of seizures occur and when is crucial for identifying suitable interictal segments. Any information that may be relevant to segment selection is therefore stored within this Seizure class.

Required fields in Seizure are **Subject ID**, **Exam ID**, and **Seizure date-time**. The latter must be unique within the exam (using **Exam ID**).

Optional Seizure fields include the Seizure duration, Type, and Onset channels. Duration determines the endpoint of the seizure, helping to know when an interictal segment can start. Further, readers may want to impose different constraints on different seizure types. For example, perhaps interictal segments should be at least 2 h away from subclinical seizures but must be at least 4 h away from focal seizures (**DP A2**). Seizure type can help with this if the information is available. Finally, Seizure onset channels might be found in clinical reports; if so, this field can help identify channels that are unsuitable for normative maps.

That concludes the written description of the database, which can be queried for various subsets of data. Below, we include the object-data-map we used for MongoDB as an example.

**End point:**

An interpretable, organised database of all subjects that can be queried to identify segments based on various criteria, in particular, those suitable for normative mapping.

```
## PIPELINE C: DATABASE CREATION

## EXAMPLE DATABASE OBJECT DATA MAPPING

from mongoengine import *

## INPUT OPTIONS FOR THE CONSTRAINED FIELDS

## General Constraints

boolean = (('0', 'FALSE'),
                  ('1', 'TRUE'))

Hemispheres = (('L', 'Left hemisphere'),
                        ('R', 'Right hemisphere'))
```

```
## Subject Constraints

Hospital_ID = (('UCLH', 'NHS UCL Hospitals'),
                        ('GOSH', 'Great Ormond Street'))

Sexes = (('M', 'born as male'),
                ('F', 'born as female'))

Handednesses = (('L', 'left-handed'),
                            ('R', 'right-handed'),
                            ('A', 'ambidextrous'))

## Exam Constraints

Exam_types = (('icEEG', 'Intracranial EEG examination'),
                         ('MRI','Magnetic resonance imaging'),
                         ('CT','Computer Tomography'))

## Treatment Constraints

Treatment_types = (('Surgery', 'Epilepsy Surgery'))

## Channel Constraints

electrode_types = (('grid', 'grid ECoG'),
                              ('strip', 'strip ECoG'),
                          ('depth', 'depth or SEEG'))

## icEEG Segment Constraints

icEEG_reference_types = (('referential', 'as recorded relative to some reference'),
                                          ('bipolar', 'some bipolar montage'))

sleep_stages = (('W', 'awake'),
                            ('N1', 'non-REM light sleep, stage 1'),
                            ('N2', 'non-REM light sleep, stage 2'),
                            ('N3', 'non-REM deep sleep'),
                            ('R', 'REM sleep, vivid dreaming'))

## Seizure Constraints

seizure_types = (('focal', 'Focal seizure'),
                       ('subclin', 'Subclinical seizure'),
                       ('sg', 'Secondary generalised seizure'))

## SETTING UP CLASSES

## Level One: Subject Class

class subject(Document):

    # required fields

    Hospital = StringField(choices=Hospital_ID, required=(True))
    subjID = StringField(required=(True), unique_with=['Hospital'])

    # optional fields
```

```
    DOB = DateField()
    Sex_at_birth = StringField(choices=Sexes, max_length=1)
    Handedness = StringField(choices=Handednesses)
    Age_of_epilepsy_onset = FloatField()

    # link to Level Two classes

    treatments = ListField(ReferenceField('treatment'))
    exams = ListField(ReferenceField('exam'))
    epilepsy_status = ListField(ReferenceField('epilepsy_status'))

## Level Two: Epilepsy Status

class epilepsy_status(Document):

    # required fields

    subject_id = ReferenceField('subject', required=(True))
    date_of_status = DateField(required=(True), unique_with=['subject_id'])

    # optional fields

    status_epilepticus = BooleanField()

    complex_partial_seizures = BooleanField()
    CPS_monthly_frequency = FloatField()

    secondary_generalised_seizures = BooleanField()
    SGS_monthly_frequency = FloatField()

## Level Two: Exams

class exam(Document):

    # required fields

    subject_id = ReferenceField('subject', required=(True))
    exam_type = StringField(choices=Exam_types, required=(True))
    exam_date = DateField(required=(True))

    # optional fields specific to exam_type = icEEG

    iceeg_report_file_location = StringField()
    raw_iceeg_folder_location = StringField()

    # optional fields specific to exam_type = MRI

    pre_implantation_mri_file_location = StringField()
    post_surgery_mri_file_location = StringField()     # if available, for resection
masks

    # optional fields specific to exam_type = CT

    post_implantation_ct_file_location = StringField()

    # link to Level Three classes

    channels = ListField(ReferenceField('channel'))
```

```
    iceeg_segment = ListField(ReferenceField('iceeg_segment'))
    seizures = ListField(ReferenceField('seizure'))

## Level Two: Treatments

class treatment(Document):

    # required fields

    subject_id = ReferenceField('subject', required=(True))
    treatment_type = StringField(choices=Treatment_types, required=(True))
    treatment_date    =    DateField(unique_with=['subject_id',    'treatment_type'],
required=(True))

    # optional fields
    outcome_ILAE = ListField(IntField())
    outcome_Engel = ListField(IntField())
    outcome_date = ListField(DateField())

    #optional fields specific to treatment_type = surgery

    surgery_type = StringField()
    surgery_pathology = StringField()
    surgery_hemisphere = StringField(choices=Hemispheres)

## Level Three: Channels

class channel(Document):

    # required fields
    subject_id = ReferenceField('subject', required=(True))
    exam_id = ReferenceField('exam', required=(True))
    channel_name = StringField(required=(True), unique_with=['exam_id'])

    # optional fields

    channel_hemisphere = StringField(choices=Hemispheres)
    electrode_type = StringField(choices=electrode_types)
    location = ListField(FloatField())
    ROI_ids = ListField(IntField())
    ROI_name = ListField(StringField())

    is_resected = BooleanField() # if within 5 mm of resection mask, DP B2
    is_spiking = BooleanField()
    is_within_structurally_abnormal_tissue = BooleanField()
    is_within_soz = BooleanField()

## Level Three: icEEG Segments

class iceeg_segment(Document):

    # required fields

    subject_id = ReferenceField('subject', required=(True))
    exam_id = ReferenceField('exam', required=(True))
    icEEG_datetime = DateTimeField(unique_with=['exam_id'], required=(True))
    icEEG_duration = FloatField(required=(True))
    icEEG_sampling_frequency = FloatField(required=(True))
```

```
    icEEG_reference = StringField(choices=icEEG_reference_types, required=(True))
    icEEG_wake_status = StringField(choices=sleep_stages)
    icEEG_segment_file_location = StringField(required=(True))

    # no optional fields

## Level Three: Seizures

class seizure(Document):

    # required fields

    subject_id = ReferenceField('subject', required=(True))
    exam_id = ReferenceField('exam', required=(True))
    seizure_datetime = DateTimeField(unique_with=['exam_id'], required=(True))

    # optional fields

    seizure_duration = FloatField()
    seizure_type = StringField(choices=seizure_types)
    seizure_onset_channels = ListField(StringField())
```

### D. Normative mapping

The **goal of pipeline D** is to use the database output to generate a regional map of normative brain activity that encompasses all interictal channel-level data.

Before starting this pipeline, some **decision points (DPs)** must be addressed. These are the reader's choice and will affect the resulting normative map. We provide the decisions made in Taylor et al. [1] where possible.

• **Decision point D1:** Will there be a constraint on the type of electrode used in the normative map (i.e., grid, depth)? Certain signal processing properties may differ across electrode types due to their physical properties or location differences. One study finds that signal characteristics appear to be similar but not identical between electrode types [39]. Nevertheless, Taylor et al. [1] included all intracranial electrode types in their normative map.

• **Decision point D2:** Will there be any further precautionary steps to identify noisy channels before constructing the map? If so, what will these be? Taylor et al. [1] located and removed channels with outlier signal range and/or variance relative to the other channels; a more detailed outline of the process can be found in **Supplementary information 1**.

• **Decision point D3:** Which preprocessing steps will be applied to the icEEG segments? Preprocessing choices can impact brain signals, including the referencing approach [40,41]. Briefly, Taylor et al. [1] applied a common average reference within a subject and bandpass filter. They also downsampled the data to a common sampling frequency across all subjects, regions, and channels and removed power line noise. Specifics can be found in **Supplementary information 2**.

• **Decision point D4:** Which metric(s) will be used for the normative map? Some metrics can be calculated directly from the icEEG segment, such as PSD and phase-amplitude coupling. The PSDs can also be used to calculate further metrics, such as relative band power (RBP), absolute band power, and power shift. Taylor et al. [1] constructed a normative map of log(RBP) in five frequency bands of interest. See **Supplementary information 3** for their steps in more detail.

• **Decision point D5:** When constructing a regional map, there may be instances where a subject has two or more channels assigned to a single ROI. How should the data from multiple channels be consolidated to the ROI level? Taylor et al. [1] took the mean of the RBP values to give each subject a single value of RBP per frequency band per implanted region.

• **Decision point D6:** How will outliers be identified at the regional level? Outliers could skew the normative distributions and can impact downstream results. For example, calculated abnormalities (pipeline E) can be drastically underestimated due to outliers artificially inflating the spread of the normative distribution. In our ongoing projects, we use an iterative leave-one-out approach, by assessing each subject against the rest of the normative map; see **Supplementary information 4** for the specific method.

• **Decision point D7:** Following all data selection steps, some subjects may have more than one segment remaining. How will multiple segments originating from the same subject be organised? This decision is dependent on the research question; the motivation for selecting multiple segments in our ongoing projects is to maximise the chance of a subject still having a segment if one (or more) was labelled as an outlier, and to establish temporal variability/stability [5]. Hence, following all data refinement, any remaining backup segments can be excluded. The reader might want to handle subjects with multiple segments differently if, for example, segments are to be used for validation or for investigating normative map variability.

Once **decision points D1–7** have been finalised, begin normative mapping (pipeline D). This pipeline is applied to *all subjects*. Alternatively, a subset can be held out for abnormality mapping (pipeline E). This pipeline assumes the reader has constructed the database outlined in pipeline C. We recommend creating a new folder to store outputs from this pipeline. See Figure 1.

1. Query the database created in pipeline C to obtain the subset of subjects that will be used for normative mapping, applying any desired constraint on electrode type (**DP D1**).
**Quality control:** Remember to consider, and be consistent with, the decisions made in temporal processing (pipeline A, **DP A1–7**).

2. Load the subset of normative mapping subjects identified in the previous step into the reader's preferred programming language.
3. Refine the data by removing all channels that are thought to represent pathology (non-normative). If available, this information was recorded in the previous pipelines. Non-normative channels are those that were labelled as:
a. Resected
b. Spiking
c. Within structural abnormalities
d. Within the SOZ

4. Refine the data further by removing any channels that are noisy or unsuitable. These are channels which:
a. Were labelled as unsuitable channels during inspection and extraction of icEEG segments.
b. Did not localise to an ROI in pipeline B/do not have mapping information available.
c. Are algorithmically detected as noisy (**DP D2**).

5. Apply icEEG preprocessing steps to all remaining channels (**DP D3**).
**Quality control:** Power line noise can differ by country. For example, data from UK hospitals has power line noise at 50 Hz, whilst data from the USA has it at 60 Hz. If removing power line noise, make sure to apply the correct filter to each batch of data.

**Quality control:** If the reader is considering larger frequency ranges, remember to apply a notch filter for harmonics where applicable.

6. Save the preprocessed segment(s) in a suitable folder location. We recommend one folder per subject, see Figure 1.
7. Calculate metrics for each channel in each preprocessed segment (**DP D4**).
8. Save the metrics calculated in the previous step in a suitable folder location. Again, we recommend one folder per subject; see Figure 1.

**Pause point:** At this point, the reader should have preprocessed interictal segment(s) for each subject, with metrics calculated for each. These segments should be stored in an easily accessible, organised manner with non-normative and unsuitable channels removed.

9. Map channel data to the regional level, with consideration of whether to/how to attain a single metric per subject per region (**DP D5**).
10. Check for and label outliers within the current normative data table **(DP D6**).
11. Reduce the normative data by removing all outliers.
12. Reduce the normative data, if necessary, to retain the desired number of segments per subject (**DP D7**).

**Quality control:** Depending on the intended research question, the reader might check the subject sample size per ROI here, or the distribution of samples, and potentially exclude any regions that are not adequate. For example, suppose the intention is to use the normative map as a baseline to calculate z-scores. In that case, each ROI should have at least 30 samples, and the distribution of samples should be approximately normal.

13. Save the resultant data table in a suitable location.

**Endpoint:** A data table that provides a regional, normative map of brain activity using the reader's desired metric(s). The map has had samples thought to represent pathology removed and has been algorithmically tested for outliers. As an example endpoint, Taylor et al. [1] took the normative distribution of RBP in a region (in a particular frequency band) to be the distribution of RBPs of all subjects with coverage in that region.

Completion of pipeline D concludes the fundamental sections of our protocol. The normative map is now ready for downstream analysis.

## E. Abnormality mapping (optional)

The **goal of this pipeline** is to compare new subjects to the normative map, with the aim of identifying abnormalities that may represent pathology.

Before starting this pipeline, some **decision points (DPs)** must be addressed. These are the reader's choice and will affect the resulting abnormality map. We provide the decisions made in Taylor et al. [1] as an example.

• **Decision point E1:** What method will be used to score subjects against the normative map? To estimate regional abnormality compared to the normative distribution, Taylor et al. [1] computed the absolute z-scores of RBP values. They then computed the maximum RBP across frequency bands for each region, thereby creating a subject-specific abnormality map. For more details, see **Supplementary information 5**. An alternative could be centile mapping, and channel-wise, as well as region-wise, abnormality mapping is also possible.

• **Decision point E2 (optional, requires resection mask):** What threshold will be implemented to define a resected region? For each subject in the abnormality cohort, Taylor et al. [1] defined a region as resected if >25% of channels within the region were removed; otherwise, regions were considered spared.

• **Decision point E3 (optional, requires resection mask):** How will the difference in abnormality scores between resected and spared regions be quantified? Abnormality measures can be computed for resected and spared regions, allowing for direct comparison. Taylor et al. [1] used the distinguishability statistic ($D_{RS}$), which is the area under the receiver operating curve. See **Supplementary information 6** for a more detailed explanation.

Additionally, some of the decisions made in pipeline D (when constructing the normative map) need to be maintained for abnormality mapping (**DP D1–5**); otherwise, the comparison between identified abnormalities and the normative baseline will not be useful. The preprocessing steps (**DP D3**) and chosen metric (**DP D4**) must be the same as in pipeline D for abnormality mapping to be viable. We recommend also keeping the electrode type (**DP D1**) and method for handling multiple channels per subject per region (**DP D5**) the same. Noisy channel detection (**DP D2**) should be applied here, but strictly to identify non-physiological channels; it should not risk excluding any genuine interictal spiking, for example.

Regional outlier detection (**DP D6**) should not be implemented, as it may identify abnormalities that are actively being sought in this pipeline. For simplicity, this pipeline assumes only one segment per subject is retained for abnormality mapping, negating the need for **DP D7**.

Once **decision points E1–3** have been finalised, begin abnormality mapping (pipeline E). Abnormality mapping is either done using a subset of subjects held out during normative mapping (pipeline D) or using new subjects. Unseen subjects will need to go through pipeline A–C before this one. This pipeline assumes the reader has the normative map created in pipeline D ready to use.

1. Query the database created in pipeline C to obtain the subset of subjects that will be used for abnormality mapping, applying any desired constraint on electrode type (**DP D1**).

**Quality control:** Remember to consider, and be consistent with, the decisions made in temporal processing (pipeline A, **DP A1–7**).

2. Load the subset of abnormality mapping subjects identified in the previous step into the reader's preferred programming language.

3. Refine the data by removing any channels that are noisy or unsuitable. These are channels which:

a. Were labelled as unsuitable channels during inspection and extraction of icEEG segments.

b. Did not localise to an ROI in pipeline B/do not have mapping information available.

c. Are algorithmically detected as noisy (**DP D2**).

*Note: We do not remove suspected pathological channels (as in step D3); the goal is no longer to obtain normative data.*

4. Apply icEEG preprocessing steps to all remaining channels (**DP D3**).

**Quality control:** Power line noise can differ by country. For example, data from UK hospitals has power line noise at 50 Hz, whilst data from the USA has it at 60 Hz. If removing power line noise, make sure to apply the correct filter to each batch of data.
**Quality control:** If the reader is considering larger frequency ranges, remember to apply a notch filter for harmonics where applicable.

5. Save the preprocessed segment(s) in a suitable folder location.
*Note: We recommend one folder per subject; see Figure 1 for the organisation of normative mapping files. Similar organisation can be implemented for abnormality mapping files.*
6. Calculate metrics for each channel in each segment (**DP D4**).
7. Save the metrics calculated in the previous step in a suitable folder location.
**Pause point:** At this point, the reader should have preprocessed interictal segment(s) for each subject, with metrics calculated for each. These segments should be stored in an easily accessible, organised manner with unsuitable channels removed. So far, we are following the same steps as pipeline D, but have not excluded non-normative data.
8. Map channel data to the regional level, with consideration of whether to/how to attain a single metric per subject per region (**DP D5**).
9. Quantify regional abnormalities through comparison to the normative map for each subject (**DP E1**).
10. [Optional] Identify which regions were resected and which regions were spared for each subject (**DP E2**).
11. [Optional] Compute abnormality measures between resected and spared regions (**DP E3**).

**End point:** A regional abnormality map for each subject in the abnormality cohort, highlighting areas that may be pathological through comparison to the normative map.

# Data analysis

This protocol aims to prepare neuroscience researchers for the data analysis stage by equipping them with the tools to construct normative maps. It does not offer advice on the data analysis itself.
However, our lab's published works provide some examples of the subsequent data analysis stage, demonstrating the type of results that can be obtained using this normative mapping protocol (or aspects of it). Below, we highlight an example result from some of our research articles. See the Methods section of each reference for details on the data analysis procedures.

**Pipelines A–E**
• Taylor et al. [1] used normative and abnormality maps to demonstrate that interictal icEEG can localise epileptogenic tissue (Figures 3 and 4).
• Kozma et al. [2] found that relative complete icEEG band power is more effective for distinguishing between surgical outcome groups than the periodic or aperiodic component alone (Figure 4).
• Horsley et al. [3] demonstrated that both connectivity and icEEG abnormalities can localise epileptogenic tissue (Figure 4), and that dMRI abnormalities could inform icEEG electrode placement (Figure 5).
• Owen et al. [4] found that a combination of magnetoencephalography and icEEG abnormalities is predictive of surgical outcome (Figure 4).
• Wang et al. [5] showed that for icEEG, the $D_{RS}$ statistic (a measure of how different surgically resected and spared tissue is) is relatively consistent over time and able to separate surgical outcome groups (Figures 2 and 3).

• Steinschneider et al. [6] used icEEG from a child with left insular ganglioglioma to identify abnormalities by comparison to a normative map (Figure 5, Table 3).

**Pipelines A–D**

• Janiukstyte et al. [7] showed that normative maps generated from scalp EEG correlate positively with those generated from icEEG and magnetoencephalography (Figure 3).

**Pipeline B**

• Wang et al. [8] showed that normalising for spatial proximity between nearby icEEG electrodes improves the prediction of post-surgery seizure outcomes (Figure 2).

• Thornton et al. [9] used long-term icEEG recordings to show that circadian and ultradian rhythms are diminished in pathological brain tissue (Figures 1 and 2).

• Gascoigne et al. [10] constructed resection masks to demonstrate that the incomplete resection of the icEEG SOZ is not associated with surgery outcomes (Figure 2).

**Pipeline B: resection masks**

• Taylor et al. [11] used pre- and post-surgery MRIs to infer that the impact of surgery leads to a $<10\%$ reduction in efficiency in the majority of subjects (Figure 4).

• Owen et al. [12] found that markers of surgical failure mechanisms, such as failure to resect magnetoencephalography abnormalities, discriminate between surgical outcome groups (Figure 5).

# Validation of protocol

This protocol (or aspects of it) has been validated in the twelve published articles outlined in the section above (**Data analysis**). The DOIs of the validation articles are also listed after the **Keywords** section.

# General notes and troubleshooting

**Glossary**

• **Channel:** One of several contact points found on an electrode, which records electrical brain activity.

• **Electrode:** A device comprising several recording channels, which is surgically implanted. Types of electrodes include depth and grid.

• **Harmonics:** Power line noise, which occurs at multiples of the initial frequency. For example, if there is power line noise at 50 Hz, it may also be present at 100 and 150 Hz.

• **Period (of data):** A subsection of icEEG data taken from a subject's full dataset, which is taken forward for further investigation/possible segment selection.

• **Pre/post-implantation:** Refers to before/after the implantation of intracranial EEG electrodes. The word operation is avoided to prevent confusion between the electrode implantation operation and the epilepsy surgery operation.

• **Pre/post-surgery:** Refers to before/after any resective epilepsy surgery. The word operation is avoided to prevent confusion between the electrode implantation operation and the epilepsy surgery operation.

• **Resected (ROIs):** ROIs that were removed during epilepsy surgery.

• **Segment (of data):** A clean, interictal icEEG segment of fixed length that is inputted into the database for use in normative (or abnormality) mapping.
• **Spared (ROIs):** ROIs that were not removed during epilepsy surgery.
• **Subject:** An individual for whom we have the necessary icEEG recordings and neuroimaging to proceed with the protocol. In our published works, this is primarily individuals with drug-resistant epilepsy undergoing evaluation for resective epilepsy surgery.

**Choice of programming environment**

As mentioned in the Software and Datasets section, the programming environment is the reader's choice. We envisage that most users of this protocol are already familiar with programming environments and/or will have a preferred environment in mind. If not, some widely used examples include MATLAB, Python, and R. Python and R are free and open source, whereas MATLAB requires a license for full usage.

Both for this protocol and in general, MATLAB, Python, and R are typically interchangeable; however, each has its pros and cons. As part of an investigation into which of the three is more suitable for teaching college students, Ozgur et al. [42] discussed the basics of each programme, along with advantages in relation to one another, which may help readers of this protocol determine which programming environment is the most appropriate to start with.

**Number of subjects**

One clear question raised when applying the full protocol is the number of subjects required to build a robust normative database. Unfortunately, there is no clear-cut minimum. The appropriate sample size is dependent on the reader's intended analysis, and as with any research, a larger number of subjects would be preferable, especially when using people with epilepsy (or any patients) to approximate normative trends.

Published work from our lab, which utilises the RAM data (https://memory.psych.upenn.edu/RAM) for normative mapping [1–7], uses a minimum of 234 subjects to construct the map.

**Decision points**

Throughout the protocol, **decision points** are included where appropriate. These are intended to be methodological choices that impact the resultant normative map and subsequent analysis. They are parameters that, when changed, lead to a different outcome, e.g., constructing a normative map of wake activity versus REM activity. Decisions that should not impact outcomes are not included. For instance, the reader's choice of programming environment used to extract interictal segments should not impact downstream results.

The inclusion of **decision points** allows the reader to tailor the protocol to their own research question. It also facilitates repeatability, allowing the protocol to be implemented for various research questions by merely changing one (or more) **decision point(s)**. Nevertheless, as a starting point for most **decision points**, we include the choices made in our primary research paper [1]. If the reader were to follow the protocol through with those decisions implemented, they would be in a position to reproduce the results in that paper.

## Supplementary information

The following supporting information can be downloaded here:
1. Supplementary information 1. Algorithmic detection of noisy channels.
2. Supplementary information 2. icEEG preprocessing steps.
3. Supplementary information 3. Calculating a normative metric: log(RBP).
4. Supplementary information 4. Identifying outliers at the regional level.
5. Supplementary information 5. Scoring new subjects against the normative map.
6. Supplementary information 6. Comparing resected and spared regions.

## Acknowledgments

We thank members of the Computational Neurology, Neuroscience & Psychiatry Lab (www.cnnp-lab.com) for discussions on the analysis and manuscript. We also thank each author on our list of published validation papers. Lastly, we thank our funders. H.W, S.J.G, and C.S are funded by the Engineering and Physical Sciences Research Council (EP/L015358/1). Y.W and P.N.T are funded by the UKRI Future Leaders Fellowships (MR/V026569/1, MR/T04294X/1). N.E is funded by Epilepsy Research Institute UK. This protocol was used in [1].

## Competing interests

The authors have no competing interests to disclose.

## Ethical considerations

Any work involving data from human subjects must go through the proper channels and attain ethical approval. This approval should be included in any subsequent research. For example, we have obtained NHS database ethics to ensure full anonymisation and databasing across NHS sites [43] separately to permission to handle and analyse anonymised data from the University Ethics Committee at Newcastle University (ref. 12721/2018). In terms of data anonymisation, care should be taken to ensure subjects with rare, identifiable epilepsy conditions, for example, cannot be identified within a cohort. Formal guidelines on databasing at scale to ensure anonymity exist elsewhere [44–46].
As previously mentioned, public datasets are also available. Some of our published works [1–7] use the RAM cohort (http://memory.psych.upenn.edu/RAM).